\documentstyle[11pt]{article}
\begin{document}
\centerline{\Large\bf Comment on "a unified scheme for flavored meson
and baryons"}

\vspace{0.6cm}

\centerline{ Yi-Bing Ding$^{a,b}$,
             Xue-Qian Li$^{c,d}$, Peng-Nian Shen$^{c,e}$}
\vspace{0.5cm}
{\small
{
\flushleft{  $~a.$ Graduate School, USTC at Beijing,
Academia Sinica, P.O.Box 3908, Beijing 100039, China}
\vspace{8pt}
\flushleft{  $~b.$ Department of Physics, University of Milan, INFN,
Via Celoria 16, 20133 Milan, Italy}
\vspace{8pt}
\flushleft{  $~c.$ CCAST(World Laboratory),P.O.Box 8730, Beijing 100080, China}
\vspace{8pt}
\flushleft{  $~d.$ Department of Physics, Nankai University, Tianjin
300071, China}
\vspace{8pt}
\flushleft{  $~e.$ Institute of High Energy Physics, Academia Sinica,
 P.O.Box 918(4), Beijing 100039, China}
}}

\date{\today}

\vspace{0.6cm}


\begin{center}

\begin{minipage}{12cm}

\centerline{\bf Abstract}
\vspace{0.2cm}

We would comment on the results of the paper "a unified scheme for flavored meson
and baryons" ( P.C.Vinodkumar , J.N.Panandya, V.M.Bannur, and S.B.Khadkikar
Eur. Phys. J. A4(1999)83 ), and point out
some  inconsistencies and mistakes in the work for
solving the Dirac equation. In terms of an example for a single
particle  we investigate the reliability of the perturbative method
for computing the Coulomb energy and  discuss the contribution to the wavefunction at origin from the Coulomb potential.
We conclude that the accuracy of their numerical
results needs to  be reconsidered.
\\
\end{minipage}
\end{center}

\vspace{1cm}

\baselineskip 22pt

In the recent paper \cite{pcv}, based on a relativistic harmonic confinement
model (RHM) with a phenomenological parameterization for the residual color
electrostatical potential and using the residual confined one-gluon-exchange
potential (COGEP) for the spin-hyperfine interaction, an attempt was made to
compose a
unified understanding of the spectroscopy of hadrons from light flavors to
heavy ones. The authors of Ref.\cite{pcv}
computed the masses of mesons and baryons of different
flavor combinations for $q\bar q, q \bar Q , Q\bar Q, qqq, qqQ, qQQ , QQQ $
etc. where $q$ and $Q$ stand for light and heavy flavor quarks respectively,
with a unique confinement strength . In this framework they obtained mass differences of
pseudoscalar-to-vector mesons as well as that for $1/2$-to-$3/2$ baryons.
They predicted the masses of the S-state flavor mesons and baryons
which are not measured yet
and evaluated leptonic decay widths of the vector mesons with no further
additional parameters. As shown in the tables of \cite{pcv}, their numerical
results are in good agreement with the experimental data, much
of which are even better than that obtained in terms of
some successful models.

The idea is obviously remarkable that one can suppose the effective mechanism
for both light and heavy quarks to be the same
and all
hadron states can be well described in a unified framework.
However, unfortunately, we find that there are some evident inconsistencies
and mistakes in their work while
solving the Dirac equation, using the perturbative
method to compute the Coulomb energy and omitting the Coulomb potential to calculate the 
wavefunction at origin. The ill-treatment greatly influences
the reliability of their numerical results.

\vspace{0.6cm}

1. On the solution of the Dirac equation.

\vspace{0.6cm}

The basic Dirac equation (3) in \cite{pcv} which describes the confined
single particle state of the quark under a mean field potential $V(r)$
is the starting point of their calculations. It reads
\begin{equation}
[i\gamma^\mu \partial \mu -M_q-V(r)]\psi_q(r)=0 \label{one}
\end{equation} 
where V(r) is a potential of form
\begin{equation}
V(r)={1 \over 2} (1+\gamma_0) A^2 r^2,
\end{equation}
which includes both scalar and vector parts.

The wavefunction of quark
$\psi_q(r)$ was expressed as a bispinor form $[ \chi_q,\phi_q]$ which
satisfies coupled equations,
\begin{equation}
[E-M_q-A^2r^2]\chi_q=-i\vec{\sigma} \cdot \nabla \phi_q \label{three}
\end{equation}
\begin{equation}
[E+M_q]\phi_q=-i\vec{\sigma}\cdot \nabla \chi_q ,\label{four}
\end{equation}
where $M_q $ is the quark mass parameter and the $E$ is the eigenvalue of 
energy.

The authors of Ref.\cite{pcv} introduced an operator $U$ of the form
\begin{equation}
U=\frac{1}{1+\frac{p^2}{(E+M_q)^2}}\left(
\begin{array}{cc}
1 & \frac{\vec{\sigma}\cdot \vec {p}}{(E+M_q)}\\
-\frac{\vec{\sigma}\cdot\vec {p}}{(E+M_q)}&1
\end{array}\right).
\end{equation}
Then $\psi_q$ can be transformed into a form where the lower
component is eliminated
\begin{equation}
U\psi_q=\left(
\begin{array}{c}
\chi_q \\0
\end{array}\right).\label{fourp}
\end{equation}
They gave  the following normalization
\begin{equation}
\langle \psi_q|\psi_q \rangle = \langle \chi_q|\chi_q \rangle=1\label{five}
\end{equation}
and the upper component $\chi_q $ satisfies
\begin{equation}
[-\nabla^2+A^2r^2 (E+M_q)]\chi_q=(E^2-M_q^2)\chi_q \label{six}.
\end{equation}

However, it is very easy to prove that U is not a unitary operator because 
\begin{equation}
U^\dagger U = U U^\dagger = \frac{1}{1+\frac{p^2}{(E+M)^2}}\neq 1,
\end{equation}
unless under the extreme non-relativistic limit.

Thus, Eq.(\ref{five}) is incorrect. Instead, we should have
\begin{eqnarray}
\langle \psi_q|\psi_q \rangle &=& \langle \psi_q|U^\dagger
(1+\frac{p^2}{(E+M^2)})U|\psi_q \rangle \nonumber \\
&=& \langle \chi_q |(1+\frac{p^2}{(E+M^2)})|\chi_q \rangle=1.\label{seven}
\end{eqnarray}
Eq.(\ref{seven}) gives the normalization condition for $\chi_q$.
In fact, we do not need the transformation operator U, because $E+M_q\neq 0$,
from Eq. \ (\ref{four}) we have
\begin{equation}
\phi_q=\frac{\vec{\sigma}\cdot \vec{p} }{(E+M)}\chi_q. \label{eight}
\end{equation}
Substituting it into Eq.(\ref{three}), a corresponding equation similar to
Eq.(\ref{six}) for $\chi_q$
can be obtained immediately. The normalization condition Eq.(\ref{seven})
is satisfied naturally. However, an important difference manifests. The lower
component of the $\psi_q$ is not zero and  cannot be neglected in the
 calculation of matrix element, which would have a great influence to the
 numerical results of Ref.\cite{pcv}.

It is not difficult to understand this inconsistency. Instead of $U$ we can
take an unitary operator $U_0=\sqrt {1+\frac{p^2}{(E+M_q)^2}}U$.i.e.
\begin{eqnarray}
U_0=\frac{1}{\sqrt{1+\frac{p^2}{(E+M_q)^2}}}\left(
\begin{array}{cc} 
1 & \frac{\vec{\sigma}\cdot \vec {p}}{(E+M_q)}\\
-\frac{\vec{\sigma}\cdot\vec {p}}{(E+M_q)}&1
\end{array}\right).
\end{eqnarray}
Using $U$ we can eliminate the lower component of  $\psi_q$ too, but
instead of the form of Eq. (\ref{fourp}) we have
\begin{eqnarray}
U_0 \psi_q=\sqrt{1+\frac{p^2}{(E+M_q)^2}}\left(
\begin{array}{c}
\psi_q^{up}\\0
\end{array}\right)\label{fourp1}.
\end{eqnarray}
Evidently, $\psi_q^{up}$ satisfies the following equation:
\begin{equation}
U_0[i\gamma^\mu \partial \mu -M_q-V(r)]U_0^\dagger \psi_q^{up}(r)=0. \label{fourp2}
\end{equation} 
Because the commutation relation between the operator $p$ and $r$ is not zero,
the equation for $\psi_q^{up}(r)$ is much more complicated than
Eq.(\ref{six}).

Actually, it is not difficult to obtain an analytical solution of  the Dirac
Eq.(\ref{one}) directly. According to the general way for solving the
Dirac equation with central potential \cite {wg}, let
\begin{eqnarray}
\psi_{q,jm}(\vec{r})=\left(
\begin{array}{c}
i\frac{G(r)}{r}\Omega_{jlm}(\frac{\vec{r}}{r})\\
 -\frac{F(r)}{r}\Omega_{jl^\prime m}(\frac{\vec{r}}{r})
\end{array}\right),
\end{eqnarray}
where $j$ is the total angular momentum quantum number, $l$ and $l^\prime$ are
the orbital angular momentum quantum numbers, $\Omega_{jlm}$ is the well-known
spherical spinor and
\begin{eqnarray}
l^\prime=2j-1=\left \{
\begin{array}{ll}
2(l+{1\over 2})-l=l+1 & for\; j=l+{1\over 2} \\
2(l-{1\over 2})-l=l-1 & for\; j=l-{1\over 2}.
\end{array} \right. \label{nine}
\end{eqnarray}
Now we insert the expression (\ref{nine}) into (\ref{one}) and can obtain the
differential equation for the radial parts G and F:
\begin{eqnarray}
\frac{dG(r)}{dr} &=& -\frac{\kappa}{r}G(r)+[E+M_q]F(r)\label{ten} \\
\frac{dF(r)}{dr} &=& \frac{\kappa}{r}F(r)-[E-M_q-A^2r^2]G(r),\label{eleven}
\end{eqnarray}
where $\kappa $ is a quantum number used frequently  in solving Dirac equation
with central potential. It is defined as
\begin{eqnarray}
\kappa=\mp(j+{1\over 2})=\left \{
\begin{array}{ll}
-(l+1) & for\; j=l+{1\over 2} \\
l & for\; j=l-{1\over 2}
\end{array}. \right. \label{twelve}
\end{eqnarray}
From Eq. (\ref{ten}) we can obtain
\begin{equation}
F=\frac{1}{E+M_q}[\frac{dG}{dr}+\frac{\kappa}{r}G].
\end{equation}
We insert the expression for $F$ into Eq.(\ref{eleven})
and obtain a differential equation for $G$ of the form
\begin{equation}
-\frac{d^2G}{dr^2}+\frac{\kappa (\kappa+1)}{r^2}G+(E+M_q)A^2r^2G=(E^2-M_q^2)G.
 \label{thirteen}
\end{equation}
It is easy to see that, if we consider $\kappa$ as an angular momentum
quantum number $l$,  Eq.(\ref{thirteen}) will be the
same as the equation for the reduced radial wave
function $u(r)=\frac{\chi_q (r)}{r}$.
Therefore, we can obtain the  solution for $ E $, $G$ and
then $F $ by using the similar method as that used by the authors of
Ref.\cite{pcv}.
The proper normalization condition is
\begin{equation}
\int_0^\infty (G^2+F^2)dr=1
\end{equation}

\vspace{0.6cm}

2. On the negative energy state

\vspace{0.6cm}

The potential under consideration is of the form of
three-dimensional harmonic
oscillator and we can easily obtain solution for eq. (\ref{six}).
The authors of ref.\cite{pcv} gave the single particle
 energy as (see Eq. (9) in \cite{pcv})
\begin{equation}
E_N=\pm \sqrt{M_q^2+(2N+3)\Omega_N(q)}. \label{fourteen}
\end{equation}
Then they claimed that "following Dirac, the negative energy state
is interpreted as antiparticle". In fact, this statement is incorrect.
Because a necessary condition for Eq.(\ref{six}) having bound-state-
solution is $E+M_q>0$ and by the definition, 
\begin{equation}
\Omega_N=A(E_N+M_q)^{1\over 2},
\end{equation}
therefore, in this case Eq.(\ref{fourteen})  only has an unique positive real solution.
Taking the negative sign in Eq. (\ref{fourteen}), we obtain  $E=-M_q$,
which is not a solution and moreover its absolute value is not equal to the
positive solution either. Thus, the negative energy solution cannot be
interpreted as one for an antiparticle.

\vspace{0.6cm}

3. On the computation of the Coulomb energy  

\vspace{0.6cm}

In \cite{pcv},  a residual Coulomb potential 
\begin{equation}
V_{coul}(q_iq_j)=\frac{\alpha_s^{eff}(\mu)}{r}
\end{equation}
was introduced and then the Coulomb part of the energy was computed
perturbatively using the confinement basis.
They gave
\begin{equation}
\epsilon_n(q_iq_j)_{coul}=\langle N|V_{coul}|N\rangle.
\end{equation}

Below we will examine the reliability of the result by this
approximate computational method. For convenience, we are not going to
repeat their calculations, instead
consider a simpler but reasonable model.
Let us put the residual Columbic potential (25)
into Eq.(\ref{six}) and explore
the reliability of the computed results in terms of the perturbative
method for a single particle state instead of the two-particle-states.
A gross estimation for the reliability which we are concerning
can be obtained accordingly.

The equation to be solved should have the form$^{\cite{kv}}$:
\begin{equation}
[-\frac{\nabla^2}{E+M_q}-\frac{\alpha_s^{eff}}{r}+A^2r^2]\chi_q^{new}=
(E-M_q)\chi_q^{new} \label{fiveteen}
\end{equation}
where ${1 \over 2}(E+M_q)$ is the dynamical effective mass of the quark.
Eq. (\ref{fiveteen}) can be rewritten as
\begin{equation}
[-\nabla^2-\frac{\lambda}{r}+A^2r^2(E+M_q)]\chi_q^{new}=(E^2-M_q^2)
\chi_q^{new} \label{sixteen}
\end{equation}
where $\lambda=(E+M_q) \alpha_s^{eff}$.

It is obvious that, if we consider $V^{per}=-\frac{\lambda}{r}$ as a
perturbative potential, the 0-th approximation of Eq. (\ref{sixteen})
is just the same as Eq.(\ref{six}). For simplicity, let us take the normalized
solution $\chi_q$ for Eq.(8) given in Ref. \cite{pcv}   as the 0-th order
approximate wave function and assume   $\Omega_0=\Omega_1=\Omega$. For the $1S$ and $2S$ states, the corresponding
0-th order approximate eigenvalues are
\begin{equation}
em_{1S}^{(0)}=E_0^2-M_q^2=3 \Omega,
\end{equation}
and 
\begin{equation}
em_{2S}^{(0)}=E_1^2-M_q^2=7 \Omega,
\end{equation}
respectively.

We can analytically derive the first-order correction to the eigenvalue
  $\langle \chi_q|V^{per}|\chi_q\rangle$ and the results are
\begin{equation}
em_{1S}^{(1)}=-(E_0+m_q)\alpha_s^{eff}(2\sqrt{\frac{\Omega}{\pi}})
\end{equation}
and 
\begin{equation}
em_{2S}^{(1)}=-(E_1+m_q) \alpha_s^{eff} (\frac{5}{3} \sqrt
{ \frac {\Omega}{\pi}}).
\end{equation}
 
Besides, we can also calculate the matrix element of the perturbative
potential $V^{per}$ in the $\{\chi_q\}$ representation, which reads
\begin{equation}
V_{12}^{per}\equiv\langle\chi_q(1S)|V^{per}|\chi_q(2S)\rangle=
-(E+M_q)\alpha_s^{eff}(\sqrt{{2 \over 3}}\sqrt{\frac{\Omega}{\pi}}).
\end{equation}

In  Ref. \cite{pcv}, the $\alpha_s^{eff}$ is a running coupling constant 
depending on both flavor and energy state and  it is
determined by a combination of several
complicated relations ( Eqs.(13)-(17) of Ref. \cite{pcv}). 
For the ground state of a meson,  a gross estimation indicates  that it is about 
$0.1 \sim 0.2 $. For convenience, we  take $\alpha_s^{eff}=0.15$ 
instead of the running coupling constant in our calculation below.

Using the data given in \cite{pcv} as
\begin{eqnarray*}
A=2166~(MeV)^{\frac{2}{3}}, M_u=M_d=82.8MeV, \\ M_s=357.5MeV,
M_c=1428MeV, M_b=4636.6MeV,
\end{eqnarray*}
 we can obtain the numerical results of $em_{1S}^{(0)}$, $em_{1S}^{(1)}$, and $V_{12}^{per}$, we list  them below in
Table 1.

\vspace{0.5cm}
\centerline{ Table 1 }
\vspace{0.3cm}
{\footnotesize
Numerical results of 0-th and 1-st approximate eigenvalues.
}
\begin{footnotesize}
\begin{center}
\begin{tabular}{|c|c|c|c|c|c|c|c|}
\hline
flavor &  $\Omega(MeV^2)$ & $\lambda (MeV)$ & $em^{(0)} (MeV^2) $ & $em^{(1)} (MeV^2)$
& $\frac{|em^{(1)}|}{em^{(0)}}$ & $V_{12}^{per}(MeV^2)$ &
$\frac{|V_{12}^{per}|}{4\Omega}$ \\
\hline
u & 46780.8 & 69.9696 &  140342 & -17076.5 & 0.121678 & -6971.44 & 0.037256 \\
\hline
s & 65992.9 & 139.242 & 197979 & -40362.1 & 0.203871 & -16477.8 & 0.062423 \\
\hline
c & 118144 & 446.270 & 354432 & -173085 & 0.488345 & -70661.5 & 0.149524 \\
\hline
b & 209335 & 1401.06  & 628005 & -723324 & 1.15178 & -295296 & 0.352659\\
\hline 
\end{tabular}
\end{center}
\end{footnotesize}

\vspace{0.6cm}

As it is well known \cite{lib}, if the approximation makes sense, the
following conditions must be respected,
\begin{equation}
|em^{(1)}| \ll em^{(0)}\label{seventeen}
\end{equation}
and
\begin{equation}
|V_{12}^{per}| \ll em_{2S}^{(0)}-em_{1S}^{(0)}=4\Omega. \label{eightteen}
\end{equation}
It is easy to check from the values in the Table 1 that 
$\frac{|em_{1S}^{(1)}|}{em_{1S}^{(0)}}$ is about $0.5$ for $c$ quark
and turns to be larger than 1 for $b$ quark. It means that the
first condition (\ref{seventeen}) is broken seriously for
$c$ and $b$ quark.
Because $\frac{|V_{12}^{per}| }{ em_{2S}^{(0)}-em_{1S}^{(0)}}$ is
about 0.35 for $b$ quark , the second
condition (\ref{eightteen}) does not hold in the $b-$case either.
The breakdown of the constraint conditions would undoubtedly
undermine the reliability of the numerical results of \cite{pcv}.

To understand this problem is not
difficult, because the Coulomb potential is a short range potential. For the
heavy quark it is more important than the confinement potential.
Our results indicate that it is not
appropriate to consider the confinement potential as the 0-th order
and the Coulomb potential as a perturbation for a heavy quark system.

\vspace{0.6cm}

4. On the wave function at the origin.

\vspace{0.6cm}

In order to compute the leptonic decay width of vector mesons,
the authors of Ref.\cite{pcv}
used the radial wave function of meson evaluated at center $R_{nS}^{q_iq_j}(0)$
 (given in eq. (26) of \cite{pcv}). It means that they only considered
 the contribution from the relativistic harmonic mean field potential and
neglected the residual Coulomb potential. However, in fact
the later is the potential
which makes the main contribution in the short range and must predominate
the value of the wave function at origin. We would like to give more
discussions on this aspect by using the simple single-particle example
described in
Eq. (\ref{sixteen}) above. The square of the wave function at origin for
$nS$ states can be obtained by using the well-known expression$^{\cite{qr}}$
\begin{equation}
|\chi_q^{new}(0)|^2=\frac{\mu}{2\pi}\langle nS|\frac{dV}{dr}|nS\rangle,\label
{eighteen}
\end{equation}
where $\mu$ is the mass of the particle. For the case corresponding
to Eq. (\ref{sixteen}), we have  $2\mu=1$. If the Coulomb potential is
omitted, the result from Eq. (\ref{eighteen}) is just
the same as that computed in
Eq.(8) of ref.\cite{pcv}. Thus, the corresponding squared wave function
at origin for 1S state is
\begin{equation}
|\chi_{q,1S}^{(0)}(0)|^2=\langle\chi_{q,1S}^{(0)}|\frac{d}{dr}
(\Omega_0^2 r^2)|\chi_{q,1S}^{(0)}\rangle=
(\frac{\Omega_0}{\pi})^{\frac{3}{2}}.
\end{equation}
Including  contributions of the Coulomb potential, by using the
0-th wave function of the Eq.(\ref{sixteen}) we can approximately estimate
the contribution  of  the residual Coulomb potential 
to the wavefunction at the origin, which reads
\begin{equation}
|\chi_{q,1S}^{(1)}(0)|^2=\langle\chi_{q,1S}^{(0)}|\frac{d}{dr}
(-\frac{\lambda}{r}) |\chi_{q,1S}^{(0)}\rangle=\lambda (\frac{\Omega}{\pi}).
\end{equation}

Still taking $\alpha^{eff}=0.15$  and in terms  of the values of Table 1
we can obtain the numerical results for $|\chi_{q,1S}^{(0)}(0)|^2$ and
$|\chi_{q,1S}^{(1)}(0)|^2$. They are shown in Table 2.

\vspace{0.5cm}
\centerline{ Table 2 }
\vspace{0.3cm}
{\footnotesize
Numerical results of 0-th and 1-st approximations for the squared wave function
at origin for $1S$ state.
}
\begin{footnotesize}
\begin{center}
\begin{tabular}{|c|c|c|}
\hline
flavor &  $|\chi_{q,1S}^{(0)}(0)|^2
( 10^6 MeV^3)$ & $|\chi_{q,1S}^{(1)}(0)|^2 ( 10^6 MeV^3)$ \\
\hline
u &   1.81709 & 1.04190\\
\hline
s & 3.04454 & 2.92494 \\
\hline
c   & 7.29277 & 16.7825 \\
 \hline
b & 17.2004 & 93.3574 \\
\hline 
\end{tabular}
\end{center}
\end{footnotesize}

\vspace{0.6cm}

The data in Table 2 clearly indicate that the contribution of the so-called residual Coulomb
potential  to the wavefunction is very close to that of the harmonic mean field for light 
flavors, moreover, it turns much larger for heavy quarks $b$ and $c$. The situation is much
more serious than for eigenvalues. Therefore, in any case, it is not plausible to neglect
the Coulomb potential at the 0-th order, even though it is a
residual one.

\vspace{0.6cm}
 
In summary, we think that the idea to construct a unified scheme for
flavored mesons and baryons is very desirable and the authors of
ref.\cite{pcv} have made a new instructive trial.
However, while solving the
basic equation established by the authors
and  calculating the residual
Coulomb energy, as well as the wave function at origin
there are some inconsistency and even mistakes, which affect the
reasonability and correctness of their numerical results and may destroy
the reliability of the solution.
Therefore, even though the motivation of Ref.\cite{pcv} is great, the adopted
method for obtaining solution
needs to be reconsidered more carefully.

\vspace{1cm}

\noindent {\bf Acknowledgments}

\vspace{0.5cm}

One of the authors (Y. B. Ding) would like to thank Prof. G. Prosperi for his
hospitality during his stay in the Department of Physics, University of
Milan. This work was partly supported by the National
Natural Science Foundation of China (NSFC) and Istituto Nazionaale di Fisica
Nuclear of Italy (INFN)


\vspace{2cm}

\begin{thebibliography}{99}
\bibitem{pcv} P.C.Vinodkumar , J.N.Panandya, V.M.Bannur, S.B.Khadkikar \
Eur. \  Phys. J. \  A. {\bf 4} 83 (1999).
\bibitem{wg} W.Greiner, \ {\em Relativiatic Quantum Mechanics  \
-- Wave Equations} (Springer-Verlag Berlin Heiderberg 1990)
\bibitem{kv} S.B.Khadkikar, K.B.Vijayakumar \ Phys. \ Lett. {\bf B \  254}
320 (1991).
\bibitem{lib} R.L.Liboff, \ {\em Introductory Quantum Mechanics},3rd edn
(Addison Wesley Longman. Inc. 1998). 
\bibitem{qr} C. Quigg and J.L. Rosener, Phys. Rep. {\bf 56} 167 (1979).
\end{thebibliography}
\end{document}